\documentstyle[preprint,eqsecnum,aps]{revtex}
\tightenlines
\begin{document}
\draft
\def \beq{\begin{equation}}
\def \eeq{\end{equation}}
\def \beqarr{\begin{eqnarray}}
\def \eeqarr{\end{eqnarray}}
\def\bc{\begin{center}}
\def\ec{\end{center}}
\def \cp{ \eta_{\sigma }(x) }
\def \cpr{ \eta_{\sigma }(\vec{r}) }
\def \ct{\cos \theta_{u}}
\def \st{\sin \theta_{u}}
\bc
\title  {$CP_N$ Solitons in Quantum Hall Systems}
\author{R. Rajaraman \cite{byline1}}

\address{School of Physical Sciences \\
Jawaharlal Nehru University\\ New Delhi 110067, \ INDIA\\ }

\date{November 2001}
\maketitle

\ec
\begin{abstract}
We will present here an elementary pedagogical introduction to $CP_N$ solitons in
quantum Hall systems. We will begin with a brief
 introduction
to both $CP_N$ models and to quantum Hall (QH) physics. Then we will focus on
spin and layer-spin degrees of freedom in QH systems and point out that these
are in fact  $CP_N$ fields for N=1 and N=3. Excitations in these degrees
of freedom will be shown to  be topologically non-trivial soliton solutions
of the corresponding $CP_N$ field equations. We will conclude with  a brief
 summary of our own recent work in this area, done with Sankalpa Ghosh.

\end{abstract}
\vskip 16pt

\section{Introduction}
$CP_N$ quantum fields  were introduced in the mid 'seventies in
  particle physics literature
  as  two-dimensional models which bore important similarities to four
  dimensional Quantum Chromodynamics. It was shown that these
  field theories were very interesting in their own right \cite{CP}. Among their
  important features was  the availability of exact solitary wave solutions
   of prototype $CP_N$ models at the classical level, even though
  the underlying field equations were coupled non-linear partial differential
  equations in 2+1 dimensions. These  solutions, obtained through elegant methods,
 could be  written in terms of simple analytic functions. They were also
  "topological solitons", i.e. they could be classified into
  homotopy sectors characterised by a winding number.
In real 4-dimensional particle physics  these beautiful
solutions remained as theoretical  discoveries in toy models with no experimental
 manifestation. However, subsequently they were shown to be physically
 realisable in an entirely different arena of physics, namely, two dimensional
quantum Hall systems.

We will present here an elementary pedagogical introduction to $CP_N$ solitons in
quantum Hall systems. We will begin with a brief
 introduction to both $CP_N$ models and to quantum Hall (QH) physics. Then we will focus on
spin and layer-spin degrees of freedom in QH systems and point out that these
are in fact  $CP_N$ fields for N=1 and N=3. Excitations in these degrees
of freedom will be shown to  be topologically non-trivial soliton solutions
of the corresponding $CP_N$ field equations. We will conclude with  a brief
 summary of our own recent work in this area, done with Sankalpa Ghosh.

\section{$CP_N$ Fields}

A $CP_N$ field is a multiplet of N+1 complex fields which are functions of some
d-dimensional space-time (which we will denote by (x)), subject to two conditions
we will list below. For the present we can leave
the space-dimensionality open and later concentrate on the case of 2 dimensions.
This field multilplet can be denoted by a $CP_N$ spinor
\beq    \eta_{\sigma}(x) \ \ \ \ \ =
 \pmatrix{  \eta_{1}(x) \cr
         \eta_{2}(x) \cr
    ... \cr
      ... \cr
            \eta_{N+1}(x)   \cr} \label{spinor} \eeq
To qualify as a $CP_N$ spinor this multiplet has to obey, \underline{at each
 point x },

(i) Normalisation : \beq   \sum_{\sigma} | \eta_{\sigma} (x) | ^2 \ \ = \ 1
 \label{norm} \eeq
and,

(ii) Equivalence under local U(1) transformations (Gauge invariance) :
 \beq \eta_{\sigma }(x) \ \ \approx \ \ \eta_{\sigma }(x)
e^{i\Lambda(x)} \label{gt} \eeq
where $\Lambda(x)$  can be a arbitrary real function of x,  but the same for all the
components $\sigma$. Altogether then there are $2(N+1) - 2  \ =  \ 2N $ real degrees of
freedom at each $x$.

The system could have any Action functional and field equations as long as
they are gauge invariant/covariant  under the U(1) tansformations above. Now,
typically, field equations involve gradients of fields. But under the gauge
transformations (\ref{gt}) gradients are not covariant.
\beq   \vec{\nabla}
\cp  \ \ \rightarrow  \ \ \ e^{i\Lambda(x)}
\big( \vec{\nabla} \cp \ + \ i(\vec{\nabla} \Lambda)\cp \big) \eeq
 However consider the "covariant derivative"
 \beq \vec{D} \cp  \ \equiv \ \ \big(  \vec{\nabla} \ + \ i \vec{A} \big)
  \ \cp \eeq
 where
\beq \vec{A} (x) \ \equiv \ i \sum (\cp)^{*} \vec {\nabla} \cp \label{A}\eeq
One can check that  $\vec{A}(x)$ is real and behaves under the  gauge transformations as
\beq \vec{A} \rightarrow \vec{A}  \ - \ \vec{\nabla} \Lambda \eeq\
Hence
\beq \vec{D} \cp \rightarrow e^{i\Lambda(x)} \vec{D} \cp \eeq

Using this, we can construct the simplest prototype $CP_N$ energy functional
for static configurations
\beq E_{pro} \ [ \cp ] \ \ = \ \  (1/2) \int d \vec{x} \sum_{\sigma}
\bigg( \vec{D} \cp^{*} \bigg) . \bigg( \vec{D} \cp \bigg) \label{action}\eeq
yielding coupled non-linear field equations
\beq \vec{D} \cdot \vec{D} \cp \ + \ \kappa \cp  \ \ = \ \ 0 \label{eqn} \eeq
where $\kappa$ is a Lagrange multiplier implementing the normalisation
condition in eq (\ref{norm}). This can be viewed as an equation for static
 (time independent) fields in some d-dimensions. Similarly, in Minkowskian d+1
dimensions, a $CP_N$ field equation would be
\beq \bigg( D_0 ^2 - \vec{D} \cdot \vec{D} \bigg) \cp \ + \ \kappa \cp  \ \
 = \ \ 0 \label{tdeqn}\eeq

where  $D_0 = \partial_t \ + \ iA_0  $.

Equations (\ref{action})and  (\ref{eqn}) are the simplest
rotationally covariant candidates for the energy functional and the
field equation
respectively for $CP_N$ systems. We may call them the prototype
$CP_N$ system. Of course any other  field equation and energy functional
for N+1 complex fields would also define a $CP_N$ system, as
long as they are covariant under the gauge transformations
(\ref{gt}) and consistent with the normalistation constraint
(\ref{norm}). Indeed the the $CP_N$ systems that appear in QH
physics do have more complicated expressions in their energy and
field equations, although they all include the basic prototype
terms above.

\section{Topological Solitons in 2 Dimensions}

Although the prototype $CP_N$ field equation eq.(\ref{eqn})
is a set of coupled nonlinear partial
diffential equations, an infinite number of exact solutions have
been obtained for them in 2 dimensions. These solutions
 are furthermore topological
solitons. We will  briefly describe them. The rest of our
discussion in this article will be limited to two space dimensions. As a first
step note that the lowest (zero) energy solutions of
eq.(\ref{eqn}) are the gauge equivalent family of spinors
 \beq  \cpr \ = \ b_{\sigma} e^{i\Lambda (\vec{r})} \label{vac} \eeq where
$b_{\sigma}$ is any constant (space independent) $CP_N $ spinor
and where the phase factor $e^{i\Lambda (\vec{r})}$ could be any
single valued function .
To see this first consider the constant solution (where $\Lambda = 0$).
Then since $\vec{\nabla} b_{\sigma} =  0 $,
the vector potential as defined in (\ref{A}) is also zero.
Hence $\vec D \ b_{\sigma} \ = \ 0 $ and $ E_{pro} [b_{\sigma}]
= 0$ . By gauge invariance of the energy, all members of the gauge class
in (\ref{vac}) will also have zero energy.

Turning to configurations of non-zero but finite energy ,
they must asymptotically (as $ r \rightarrow \infty $) tend to this zero
energy solution :
\beq \cpr \rightarrow \ b_{\sigma} e^{i\Lambda (\theta)}
\label{asym}\eeq
 where $\theta$ is the angular coordinate on the plane. Note that the angular gradient of
such configurations behaves asymptotically as
\beq \nabla_{\theta} \cpr \ \rightarrow \  {1 \over r}
\ \partial_{\theta} \Lambda\  \cpr \eeq
which is not square integrable. But it is not the plain gradient which occurs
in the energy functional (\ref{action}), but the covariant gradient which does
vanish  sufficiently fast asymptotically for the energy integral to have a finite
value.

Thus any finite energy configuration corresponds to a particular function
$\Lambda (\theta)$ on the circle at spatial infinity. This function is clearly
a mapping of a circle into a circle and can be classified by a winding
number ( the first homotopy group $\Pi_{1} [S_1] $ is the group
of integers). An explicit expression for this winding number in
terms of the asymptotic behavior (\ref{asym}) is
\beqarr n \ &=& \ {1 \over 2\pi}\int d\theta {d\Lambda  \over d\theta}
\nonumber \\
&=& {1 \over 2\pi}\int d^{2} r \epsilon_{\mu \nu} (D_{\mu}
\cp)^{*} \ (D_{\nu} \cp) \eeqarr

Exact solutions are available analytically in every topological sector (i.e. in each class
of configurations characterised by a given value of the winding number.)
It can be derived that
\beq \eta_{\sigma}(z) \ = \  \ K(z) \ \
\pmatrix{ 1 \cr
 w_2(z) \cr
         w_{3}(z) \cr
    ... \cr
      ... \cr
            w_{N+1}(z)   \cr} \label{spinor} \eeq
is an exact solution of the field equation (\ref{eqn}), where $z = x + iy$,
 $w_{\sigma}(z)$ are any analytic functions of z and K(z) is the
 normalisation factor.  For example , it can be checked that

  \beq \eta_{\sigma}(z) \ = \  \ {1 \over \sqrt{a^2 + N r^{2n}}} \ \
\pmatrix{ a \cr
 z^n \cr
         z^n \cr
    ... \cr
      ... \cr
            z^n   \cr} \eeq
 where a is any constant, is an exact solution. As $r \rightarrow \infty$,
  it behaves as
 \beq  {1 \over \sqrt{N}} \ e^{i n\theta}
\pmatrix{ 0\cr
 1 \cr
   1 \cr
    ... \cr
      ... \cr
            1 \cr}        \eeq
            and clearly has a winding number $n$ in its phase.

These exact solutions are for the prototype $CP_N$ model (\ref{eqn}).
A realistic physical system describable by a $CP_N$ field will in general have
a more complicated energy functional and field equation. But for most such
physical systems such as those which appear in the quantum Hall phenomena
the lowest energy solution is still a space-independent spinor.
Therefore localised finite energy solitons will still obey the asymptotic
condition (\ref{asym}) and be characterised by
the same winding number. Lastly, while we have presented here only static
solutions in 2 space dimensions, time dependent moving solitons can be
 obtained for eq(\ref{tdeqn}) by boosting.

 For a more detailed review of $CP_N$ solitons see \cite{Raj}.

\section{Quantum Hall Systems}

Since our subject deals with $CP_N$ solitons in Quantum Hall (QH)
systems, we would like to give some sort of an overall introduction to
the latter for those who may need it.  This is a vast subject. Further,
the basic phenomena referred to as the Quantum Hall Effect (QH) are
widely known. Therefore even though we will begin from the beginning
our overview will be channelised to focus  only on those of aspects this
system which form pre-requisites to undestanding its $CP_N$ excitations.

Recall the classical Hall problem of electrons moving in the x-y plane
confined in the $\hat{y}$ direction by boundaries, and in the
presence  of crossed electric and magnetic fields $\vec{E} = E_x \hat{x}$
and $\vec{B} = -B \hat{z}$ respectively . As the electrons begin to move in
the x-direction because of $E_x$, the Lorentz force due to the
magnetic field will push the electrons towards the y-boundary where they accumulate
and produce  a transverse  electric field $E_y \hat{y}$
which eventually balances the Lorentz force vB/c. As a result the electrons end up
moving purely along the $\hat{x}$ direction although the total electric field is
$\vec{E}=E_x \hat{x} + E_y \hat{y}$.  The electric current can be written as $\vec{j} =
\underline{\sigma}\cdot\vec{E}$ where $\underline{\sigma}$ is the
conductivity matrix which can be easily calculated using the Drude formulae.
 Its  diagonal elements are $\sigma_{x x} \
= \ \sigma_{y y} \ = \ n e^2 \tau / \mu$  (where $\mu$ = electron mass,
$\tau $ is the collision time and
$n$ is the electron density) . Its  off-diagonal
element, the Hall conductivity, is given by $\sigma_{x y}=nec/B $.
In terms of the "filling factor" $\nu$
defined as the ratio of the density of electrons to fluxons,
\beq \nu \equiv {n \over B / \phi_0} \label{nu} \eeq
where $\phi_0 = hc/e$ is the
unit of flux, the Hall conductivity can be written as
$\sigma_{x y} = ( e^{2}/ h ) \nu $.
These expressions for the conductivity tensor were obtained for the
simplest possible situation, that of  non-interacting classical
planar electrons  in a perpendicular magnetic field. Electrons in real
macroscopic experimental samples are  much more complicated. They interact with one another,
with the ions in their environment, and obey the rules of many-body
quantum mechanics. One would expect the behaviour of their conductivity
to be in general quite complicated and messy, as compared to
the simple results above. But when Hall effect experiments were done on
exceptionally pure samples of 2D electron gas at very low
temperatures and very high magnetic fields, it was found that the Hall
conductivity $\sigma_{x y}$ as a function of the filling fraction
$\nu$ revealed a
startlingly simple pattern. It contained, as a function of $\nu$ ,
a series of extraordinarily flat plateaus, with a flatness accurate to better than 1
in $10^7$ .  These plateaus were first found to
occur at integer filling fractions with Hall conductance values quantized to
be the same integer in units of $e^2/h$.  Furthermore, at those filling
fractions where $\sigma_{xy}$ had plateaus, the diagonal resistivity
$\rho_{x x}$ was found to be zero.  These phenomena were called the Integer QH Effect
(IQHE). Subsequently, in experiments involving higher magnetic fields
and higher mobility samples, the same phenomenon of plateaus in
$\sigma_{xy}$ and of vanishing of $\rho_{xx}$ was also found at
fractional values of the filling factor. These fractions (with one
exception, still being understood) corresponded to odd integers in
their denominator. This is often called the fractional QHE (FQHE).

We will concentrate here on the particular case of $\nu = 1$ where the
$CP_N$ solitons of interest to us appear. Fortunately, this is also
the value of $\nu$ at which the physics of QH effect is most easily
understood. Let us again start with non interacting
electrons in a transverse uniform B field, but now treat it  quantum
mechanically and worry about the effect of interactions  later.
This problem, solved fully and exactly long ago by Landau, is a now a
standard textbook problem in quantum mechanics (see for example \cite{mappy}).
The results in brief are as follows. The system can be
mapped into a pair of harmonic oscillators, one with frequency zero and
the other with $\omega = \omega_c \equiv eB/ \mu c $. Excitations of
the latter lead to energy levels
 \beq  E_n \ = \ (n+ 1/2) \hbar \omega_c \eeq
These are the famous Landau levels. Each level is highly
degenerate, corresponding to
excitations of the other oscillator which has zero-frequency .
The degenerate states lying in the lowest Landau level (LLL)
have wavefunctions (in the symmetric gauge) of the form
\beq \phi_m \ = \ z^{m} e^{- |z|^{2}/4l^2} \label{states} \eeq
where z=x+iy,  $l^2  = \hbar c/eB$
 and the integer $m$ ranges from zero to infinity.

The degeneracy is formally infinite for an infinite plane, but on a finite sample
it can be shown to be equal to the number of fluxons
$N_{deg} = B.A / \phi_0 $ where A is the area of the sample. Therefore
when the filling fraction as defined in (\ref{nu}) is unity, the total
number N of electrons
exactly equals the number of states in the LLL. Consequently, at unit
filling, the ground state of the system will correspond to occupying
all the states of the LLL and leaving all higher Landau levels empty.
 The system clearly has an energy gap
equal to $\hbar \omega_c = \hbar eB/\mu c$, which is very large for
large magnetic fields. [ It is this large gap and the associated
incompressibility that is responsible for the  occurance of the
Hall plateaus at $\nu=1$.  But we cannot afford to present that explanation
 here since we have to rapidly progress towards $CP_N$ solitons.]

Next consider the wavefunction of that $\nu=1$ many-body ground state ,
still staying within the non-interacting approximation. It will be a
Slater determinant of all the one-electron states in the LLL given in
(\ref{states}). Apart from the gaussian factor in each state, this is a
determinant of polynomials in z, which is just the van der Mont
determinant and can be rewritten in the Jastrow form. Hence
\beq \Psi_{\nu=1} \ = \ \Pi _{i<j}  \ (z_{i} - z_{j})
 \ \ exp \big( - \sum_{i} |z_i|^{2} / 4l^2 \big) \label{Laugh}\eeq
This is the famous Laughlin wavefunction for the $\nu = 1$ ground
state. We have derived it only in the non-interacting approximation.
But Laughlin proposed that this wavefunction will be a very good
approximation even when interactions of the electrons with one another
and with impurities are taken into account. The reason for this is that
this wavefunction already carries many of the desired features of the
exact wavefunction. It is antisymmetric, as required by the Pauli
principle. It is an eigenfunction of total angular momentum, as befits
the ground state of a system which is circularly symmetric. It vanishes
whenever two electrons coincide ---  a feature that will  reduce all pairwise
Coulomb energies. Finally, in the presence of interactions, one would
expect the ground state to contain some admixture also of states fronm
the higher Landau levels. But in the limit of very large magnetic
fields, the energy gap $e\hbar B/\mu c $ is so large that such
admixture will be small. These arguments suggest that the Laughlin
wavefunction will be sturdy even in the presence of interactions.
Indeed the Laughlin wavefunction has been found to be in excellent
agreement with numerical calculations.

\section{Spin Excitations}

The arguments in  the preceding section were incomplete in that the spin
degrees of freedom were not considered in the discussion. Even though
the electrons are treated as two-dimensional with respect to their
 coordinates, they are physical 3-dimensional electrons and do carry spin.
The spin part of the wavefunction has to be specified. Now, Pauli
principle requires antisymmetry  of the entire wavefunction
including spin. But the Laughlin wavefunction (\ref{Laugh}), which seems
to be a very accurate approximation to the correct wavefunction,
is already antisymmetric in coordinates $z_i$. Therefore the spin part
, suppressed in eq(\ref{Laugh}), must be fully symmetric. That is , all
the electron spins must be polarised in the same direction. Thus the QH
ground state at $\nu=1$ is a ferromagnet for the same reason that
usual magnets are , viz. to minimise the exchange Coulomb energy.
Given that there is a magnetic field along the z-direction one expects
the polarisation to be in the same direction.

Although the magnetic field is very strong, its coupling to the spins
is not prohibitively large because of the  effective g-factor for
electrons is reduced in the layer sandwiched between the two
semiconductors . As compared to the value of 2 for free electrons in vacuum
it can be as low as 0.4 here. Of course that is enough to allign all
spins along the B field in the ground state, but excited states are possible
at reasonably low energy where some spins point away from the z direction.
Indeed one would expect that the low energy excitations of the system
can be described solely by various spin textures , with the coordinate part of
the wavefunction still remaining in the LLL since  any admixture with
 higher Landau levels will cost heavily.

Thus the low energy dynamics of the $\nu=1$ system can be studied by
going to the continuum limit and treating the system as a
two-dimensional field of unit vectors $\vec{m}(\vec{x})$ at each point,
describing the direction of the spin at that point. This field of unit
vectors has a long history under the name of the Non-linear O(3) model.
But it is also just a $CP_1$ field (see \cite{Raj}). Given a general $CP_1$
spinor denoted by $ \cpr \ \  = \pmatrix{ \alpha(\vec{r}) \cr
\beta(\vec{r}) \cr} $ the quantity $\vec{m} \equiv \langle
\eta | \vec{\sigma}  |  \eta \rangle $  , where $\sigma_i$ are Pauli matrices,
will be a unit vector. Thus a $CP_1$ field is also a unit vector field.
The homotopy classification for $CP_N$ discussed in sec II can also be
recast , for N=1, in terms of the unit vector field $\vec{m}(\vec{r})$. The
boundary condition (\ref{asym}) on $\cpr$ corresponds to having the unit
vector $\vec{m}$ take the ${\underline same}$ value everywhere on the
 boundary of
two-dimensional space , which can be therefore compactified into a
2-sphere.  Hence any such field configuration $\vec{m}(\vec{r})$ is a mapping
of the 2-sphere $S_2$ in coordinate space into the 2-sphere of  spin
directions. These mappings $S_2 \rightarrow S_2$  are again classified by a
winding number.  These configurations are two-dimensional analogues of four-dimensional
configurations studied long ago by Skyrme (see\cite{Raj} for
references), and are called Skyrmions.

As we have already mentioned exact solutions in all topological classes
are available for prototype $CP_N$ models in two space dimensions for
all N. In the case of the nonlinear O(3) model, its Skyrmion
solutions had already been discovered by Belavin and Polyakov
before  its generalisation to $CP_N$ models had been  developed \cite
{Poly}. Turning to QH systems Sondhi et al \cite{Sondhi}, in
 a very intersting paper, showed
that not only can these exotic Skyrmion excitations occur at
$\nu=1$  , but that they are in fact the lowest energy
excitations, lower than single spin-flips. Subsequently experimental
support for the existence of such Skyrmionic ($CP_1$) excitations was
also found \cite{Barrett}.

\section{Psuedo(layer)spin in QH systems}

Following the spectacular quantum Hall results for electrons in a
layer, more complicated experiments were done using samples that
contained two parallel layers of electrons \cite{Eisen}. Some
more interesting results emerged.  One would expect of course to
see results where the system behaves as a simple additive sum of
each layer. Thus one would expect to see quantum Hall plateaus
at total filling, in both layers together, of $\nu_{total} = 2$
or $\nu_{total} = 2/3$  corresponding to the observed single
layer plateaus at $\nu$ equal to 1 and 1/3 respectively. Indeed this
is what is seen when the layer separation $d$ is large.  But when $d$
is reduced to about $3l$ quantum Hall plateaus appear at
    \underline{total} filling $\nu_{total}$   equal to unity.
The elecrostatic capacitance energy between the two layers would require  them
to have equal densities of electrons which corresponds to
a filling of 1/2 in each layer . But there is no quantum
Hall effect is seen in mono-layers at $\nu = 1/2$ .

Therefore this plateau at $\nu_{total}$ = 1 clearly cannot be
understood by thinking of the system as a pair of independent
layers. Rather, the phenomenon must reflect  some sort of a
quantum coherence between the two layers.  An ingenious formulation for
 understanding this, developed by Girvin , Macdonald and co-workers is to
associate a normalised 2-component layer-spin or pseudo spin
$\pmatrix{ \alpha \cr \beta \cr} $ to each electron
\cite{Girvmac} .  These components $\alpha$ and $\beta$ give the
amplitude for the electron being in the upper and lower layers
respectively. Let us for the moment suppress real spin and see
what the occurance of the plateau means for the pseudospin. As
we mentioned earlier, a $\nu=1$ Hall plateau is well described by the
antisymmetric Laughlin wavefunction (\ref{Laugh}). If this
should also hold for the double layer, then the Pauli principle requires that
the pseudospin of the electrons should be fully symmetric, i.e. the system
must be a pseudospin ferromagnet similar to (and for the same
exchange energy reducing reasons) the real spin  ferromagnetism.
Further, if there is even the smallest tunnelling
probability between the two layers, the ground state will be a
symmetric superposition of the two layers, ie. be in the pseudospin state
$  \pmatrix{  1 / \sqrt{2} \cr  1 / \sqrt{2} \cr}$. In other words the
pseudospin magnet will
point along the x-direction. This again is similar to the real spin
magnet pointing along the z-axis because of its Zeeman coupling to the
magnetic field. The
tunnelling term in the Hamiltonian will act as the analogue of
the Zeeman coupling for the pseudospin.

With the ground state being a ferromagnet in layer space, once again
the system will carry low energy excitations corresponding to different
peudospin textures. Asymptotically they will have to go to the ground state
value, along the x-direction, but in the interior have any smooth
configuration of direction vectors. Once again we have an O(3) or $CP_1$
field in compactified 2-space, giving rise to a topological classification
of all solutions by a winding number. Such topological solutions
called bi-merons in this context, were first discussed in detail by Moon et al
\cite{Moon}. S.Ghosh and I also studied these solutions and evaluated
their detailed profiles and energies \cite{Ghosh1}.

\section{Spin-Pseudospin Intertwined $CP_3$ Solitons}

Now let us add on spin degrees of freedom to the preceding discussion
of double layer systems. The full Hall fluid ground state
 at $\nu_{Total}=1$  will be  ferromagnetic in both spin and pseudospin,
 with a coordinate dependence given by the Laughlin wavefunction.
The combined spin and pseudospin part of the wavefunction can be described
by a 4-component multiplet:
\beq    \eta_{\sigma}(x) \ \ \ \ \ =
 \pmatrix{  \eta_{1}(x) \cr
\eta_{2}(x) \cr
 \eta_{3}(x) \cr
  \eta_{4}(x)   \cr} \label{spinor} \eeq
where the spin-pseudospin index $\sigma = 1,2,3,4$ corresponds
to amplitudes that the electron is in the upper-layer up-spin,
upper-layer down- spin, lower layer up-spin and lower -layer
down-spin states respectively. Such 4-components spinors were first studied in
QH systems by Arovas et al.\cite{Arov}and by  Ezawa \cite{Ezawa}. Since these
probabilites must add up to one, the spinor has to be  normalised and looks
like a $CP_3$ spinor. But that requires the further restriction that
 the spinor  be defined only modulo a local gauge
transformation common to all four components. This in turn requires that the
energy functional of the spinor field enjoy a corresponding gauge invariance.
Ghosh and I \cite{Ghosh3} ensured that this was so by calculating the energy of
this 4-component field starting from the microscopic Hamiltonian, following the
procedure used by the Indiana group \cite{Girvmac},\cite{Moon} for the
purely pseudospin case. Let us summarise how this is done.
Let us work in the second quantised formalism in terms of the 4-component electron field
$ \psi^{\dag}_{\sigma}(\vec r) $.

The microscopic Hamiltonian is
\beqarr H \ &=& \   \ \sum_{\sigma,\delta } \int d\vec r
\psi^{\dag}_{\sigma}(\vec r) \ \big( \tilde{g} \hat{\sigma}_{z} \ - \ t
\hat{\tau}_{x} \big)
_{\sigma \delta} \ \psi_{\delta}(\vec r)  \nonumber \\
  &+& \  \frac{1}{2} \sum_{\sigma_{1},\sigma_{2} =1}^{4}
\int d \vec r_{1}
d \vec r_{2}\psi^{\dag}_{\sigma_{1}}(\vec r_{1})\psi^{\dag}_{\sigma_{2}}
(\vec r_{2})V^{\sigma_{1} \sigma_{2}}
(\vec r_{1} -\vec r_{2})\psi_{\sigma_{2}}(\vec r_{2})\psi_
{\sigma_{1}}(\vec r_{1}) \label{H} \eeqarr
 In the above , the
Coulomb potential $ V^{\sigma_{1} \sigma_{2}}$ depends on
whether the particles are in the same layer or different layers,
 $\hat{\sigma}_{z}$ and $\hat{\tau}_{x}$ are  spin and pseudospin
matrices suitably generalised as $4 \times 4$ matrices on the outer
product space of spin and pseudospin, and $\tilde{g}$ and $t$ are the
Zeeman and tunnelling couplings.
[The kinetic term (in the presence of the B field ) can be suppressed for our
purposes since our excitations involve only LLL states
all of which carry the same constant energy of $ \hbar \omega_{c} /2$. ]
The field theoretic state vector corresponding to any given spin-pseudospin
texture
$\eta $ can be written as
\beq \mid \Psi \rangle \ = \ \prod_{X} \ [ \sum_{\sigma} \
C^{\dagger}_{\sigma X}  \eta_{\sigma}(X) ] \ \mid 0 \rangle  \label{state} \eeq
where $\mid 0 \rangle$ is the vacuum (no electron) state , $X$ stands
for Landau gauge orbitals  and $\eta_{\sigma}(X)$ is an orbital
dependent 4-spinor. The energy functional for a given spin-pseudospin texture
 is then
obtained to leading order by evaluating the expectation value of
 the Hamiltonian  (\ref{H}) in the state (\ref{state}). The result,
 upon following the same steps as pionerered by the Indiana group
 \cite{Girvmac}, \cite{Moon} is
\beqarr  E [a_{\sigma}]
& = &
  \frac{1}{2 \pi l^{2}} \ \int d{\bf r} \bigg[  \tilde{g}  \
\bigg(|a_{1}|^{2} - |a_{2}|^{2} +
 |a_{3}|^{2} - |a_{4}|^{2} \bigg)
 - \ t \bigg(a_{1} a_{3}^{*} +
 a_{2} a_{4}^{*} \ +  \ h.c. \ \bigg) \bigg] \nonumber \\
  &+& \beta
\int d{\bf r}( |a_{1}(X)|^{2} + |a_{2}(X)|^{2}  \ - \
          |a_{3}(X)|^{2}  -  |a_{4}(X)|^{2}   )^{2} \nonumber \\
 &+&
2\rho^s  \int d{\bf r} \bigg[
\sum_{i=1,4}(\partial_{\mu}a^{i\ast}(\vec r)\partial^{\mu}a^{i}(\vec r))
+(\sum_{i=1,4}a^{i \ast}(\vec r)\partial_{\mu}a^{i}(\vec r)^{2}
\bigg] \nonumber \\
&+& (\rho^{d} - \rho^{s})
 \int d{\bf r}
 \bigg[a^{1}a^{3 \ast} \vec \nabla^{2} (a^{3}a^{1 \ast})
+ a^{1}a^{4 \ast} \vec \nabla^{2} (a^{4}a^{1 \ast}) \nonumber \\
                                                        &     & \mbox{}   +
a^{2}a^{3 \ast} \vec \nabla^{2} (a^{3}a^{2 \ast}) +
a^{2}a^{4 \ast} \vec \nabla^{2} (a^{4}a^{2 \ast})  \  +  \ h. c. \bigg]
\label{INTE} \eeqarr

where the constants $\beta , \rho^d$ and $\rho^s$ are  calculated from
 the direct and exchange Coulomb energies.

Note that the third term in the energy functional (\ref{INTE}) is just the
protoype $CP_3$ energy in (\ref{action}). Our full expression for the energy is more
complicated. It can however be noticed that all the other terms are also gauge
invariant under the U(1) transformations (\ref{gt}). Therefore we $\it{are}$ dealing with
a $CP_3$ theory. All the general discussion given earlier for $CP_N$ theories apply.
Topological Soliton solutions can be  obtained for the field equations which in turn
can be  derived by extremising the energy (\ref{INTE}).

Explicit Soliton solutions have been obtained by Ghosh and I, by
numerically solving the coupled non-linear partial differential
equations that  arise when (\ref{INTE}) is extremised. In
particular we concentrated on interesting new topological $CP_3$
Solitons where the spin and pseoduspin intertwine non-trivially.
For example the very simple texture
\beq   A \pmatrix{
               \lambda \cr
         z-b \cr
	0 \cr
            z+b   \cr} \label{sub} \eeq

corresponds to a spin-Skyrmion in the upper layer and also a "bi-meron" in the
layer spin of the downspin component. This simple ansatz will of course not satisfy
the full field equations. But we have obtained numerical solutions with similar
intertwineed spin-pseudospin topology. Lack of space here does not allow us to
describe in detail these solutions . Readers interested in their detailed profile
as well as the numerical methods used are referred to reference(\cite{Ghosh3}).

We have also calculated their energy and
minimised it with respect to parameters in the ansatz. The resulting cost
of creating a pair of such topologically intertwined spin-pseudospin
excitations comes out to be about $1.2 (e^{2} / \epsilon l)$ as compared
 to particle hole excitations which cost about $1.25 (e^{2} / \epsilon l)$ .
That the former energy is a little smaller not be taken  seriously given the
various approximations that have gone into our energy calculations. All one
can say is that it is possible
that our topological $CP_3$ excitations may well be the lowest in energy, but
to be sure of this one must make longer and more precise calculations.

\section{Acknowledgements}

It is a pleasure to thank the organisers of the GIN 2001 conference,
especially Professor Radha Balakrishnan and Professor Vladimir Gerdjikov,
for their kind invitation and very warm hospitality .

\end{document}